\documentclass[twocolumn]{aastex63}

\usepackage{xcolor}

\definecolor{my_color}{HTML}{3a18b1}

\newcommand{\teff}{\mbox{$T_{\rm eff}$}}
\newcommand{\rbirth}{\mbox{$R_{\rm birth}$}}
\newcommand{\rcurrent}{\mbox{$R_{\rm current}$}}

\newcommand{\feh}{\mbox{$\rm [Fe/H]$}}
\newcommand{\mg}{\mbox{$\rm [Mg/Fe]$}}

\newcommand{\alphafe}{\mbox{$\rm [\alpha/Fe]$}}

\newcommand{\logg}{\mbox{$\log g$}}

\newcommand{\TSun}{\ifmmode {T_{\odot}}\else${T_{\odot}}$\fi}
\newcommand{\LSun}{\ifmmode {\logg_{\odot}}\else${\logg_{\odot}}$\fi}
\newcommand{\FSun}{\ifmmode {\feh_{\odot}}\else${\feh_{\odot}}$\fi}
\newcommand{\mSun}{\ifmmode {M_{\odot}}\else${M_{\odot}}$\fi}

\newcommand{\god}{6,078}
\newcommand{\gud}{262}
\newcommand{\aod}{5,832}
\newcommand{\aud}{224}
\newcommand{\nod}{5,814}
\newcommand{\nud}{5,340}

\usepackage{appendix}
\usepackage{amsmath}
\usepackage{mathtools}

\shorttitle{Metallicity Trends in Galactic-Scale Planet Occurrence}
\shortauthors{Rampalli et al.}

\begin{document}

\title{Disentangling Metallicity Effects in Hot Jupiter Occurrence Across Galactic Birth Radius and Phase-Space Density}

\correspondingauthor{Rayna Rampalli}
\email{raynarampalli@gmail.com}

\author[0000-0001-7337-5936]{Rayna Rampalli}
\altaffiliation{NSF GRFP Fellow} 
\affiliation{Department of Physics and Astronomy, Dartmouth College, Hanover, NH 03755, USA}

\author[0000-0001-5082-6693]{Melissa K. Ness}
\affiliation{Research School of Astronomy \& Astrophysics, Australian National University, Canberra, ACT 2611, Australia}

\author[0000-0003-4150-841X]{Elisabeth R. Newton}
\affiliation{Department of Physics and Astronomy, Dartmouth College, Hanover, NH 03755, USA}

\author[0000-0001-7246-5438]{Andrew Vanderburg}
\affiliation{Department of Physics and Kavli Institute for Astrophysics and Space Research, Massachusetts Institute of Technology, Cambridge, MA 02139, USA}

 \author[0000-0003-2027-399X]{Tobias Buck}
 \affiliation{Universität Heidelberg, Interdisziplinäres Zentrum für Wissenschaftliches Rechnen, Im Neuenheimer Feld 205, D-69120 Heidelberg, Germany}
 \affiliation{Universität Heidelberg, Zentrum für Astronomie, Institut für Theoretische Astrophysik, Albert-Ueberle-Strae 2, D-69120 Heidelberg, Germany}

\author[0009-0002-6271-1652]{Jessica Mills}
\affiliation{Research School of Astronomy \& Astrophysics, Australian National University, Canberra, ACT 2611, Australia}





\begin{abstract}
 
We explore how the correlation between host star metallicity and giant planets shapes hot Jupiter occurrence as a function of Galactic birth radius (\rbirth) and phase-space density in the Milky Way disk. Using the GALAH and APOGEE surveys and a galaxy from the NIHAO simulation suite, we inject hot Jupiters around stars based on metallicity power laws, reflecting the trend that giant planets preferentially form around metal-rich stars. For \rbirth\ $\geq 5$ kpc, hot Jupiter occurrence decreases with \rbirth\ by $\sim -0.1\%$ per kpc; this is driven by the Galaxy’s chemical evolution, where the inner regions of the disk are more metal-rich. Differences in GALAH occurrence rates versus APOGEE's and the simulation's at \rbirth\ $< 5$ kpc arise from survey selection effects. APOGEE and the NIHAO simulation have more high-$\alpha$ sequence stars than GALAH resulting in average differences in metallicity (0.2--0.4 dex), $\alpha$-process element enrichment (0.2 dex), and vertical velocities (7--14 km/s) at each \rbirth\ bin. Additionally, we replicate the result of \cite{Winter20}, which showed that over 92\% of hot Jupiters are associated with stars in phase-space overdensities, or ``clustered environments." However, our findings suggest that this clustering effect is primarily driven by chemical and kinematic differences between low and high-$\alpha$ sequence star properties. Our results support stellar characteristics, particularly metallicity, being the primary drivers of hot Jupiter formation, which serves as the “null hypothesis” for interpreting planet demographics. This underscores the need to disentangle planetary and stellar
properties from Galactic-scale effects in future planet demographics studies.


\end{abstract}

\submitjournal{AJ}
\accepted{June 17, 2025}
\section{Introduction}\label{sec:intro}
The wealth of complementary astrometric, spectroscopic, and photometric survey data now provides a unique opportunity to explore planet demographics on a Galactic scale. Exoplanet missions such as the Transiting Exoplanet Survey Satellite \citep{Ricker15} and Kepler \citep{borucki2010} have discovered over 5,000 exoplanets, while kinematic and abundance measurements for millions of stars from surveys like Gaia \citep{GaiaMission} have revolutionized the field of Galactic archaeology. With upcoming missions like the Nancy Roman Space Telescope \citep{Spergel15} and PLAnetary Transits and Oscillations of Stars (PLATO, \citealt{platomission}), we will soon probe even more regions of the Galaxy for exoplanets, including the Galactic bulge and thick disk. The combination of these diverse datasets enables us to connect planet demographics to the Galactic context and investigate how planet formation varies across different Galactic environments.

\subsection{Galactic-scale Influences on Planet Occurrence and the Caveats}
\cite{Ballard24} discusses the merits of considering Galactic-scale impacts on exoplanets. Particularly, in this context, we move beyond the isolated star-disk systems that have been traditionally used to understand planet formation (e.g., \citealt{Armitage11,Williams11,Winn15}). Several works consider planet occurrence as a function of Galactic-scale parameters such as phase-space stellar density, stellar relative velocity, Galactic location, and stellar Galactic oscillation amplitude (e.g., \citealt{Winter20,Dai21,Boley21,bashi22,Zink23}), which have been complemented by various simulation and modeling results (e.g., \citealt{Bitsch20, Nielsen23,boettner24,hallatt24}). 

Considering the Galactic context requires considering the inherent trends that exist between stellar ages, chemical abundances, and kinematic distributions across various Galactic environments. The two main components of the Milky way disk are the thick and thin disks, also known as the high-$\alpha$ and low-$\alpha$ sequences, that are of different scale heights and formed at different times \citep{Gilmore83,Gratton96,Bensby2003,Adibekyan13,RecioBlanco2014}. Stars in the high-$\alpha$ sequence are older in age, have higher velocities (often described as kinematically ``hot"), are lower in iron metallicity (hereafter referred to as metallicity for brevity), and are enriched in $\alpha$-process elements from core-collapse supernovae (indicative of rapid star formation) compared to younger, kinematically ``cool", metal-rich, and $\alpha$-depleted stars in the low-$\alpha$ sequence. 
Stellar age-velocity and age-metallicity relations, along with established planet frequency-metallicity and planet frequency-age correlations \citep[e.g.,][]{Gonzalez97,Santos2004,Fischer05,Johnson10, Mortier13,Guo17,Adibekyan19,OsborneBayliss20,MiyazakiMasuda23,Chen23}, are thus linked to the evolutionary history of the Galaxy. 


In addition to the vertical trends with stellar metallicity and velocity, disk stars born near the Galactic center are more metal-rich and $\alpha$-enriched from earlier, rapid star formation \citep{Bland-Hawthorn16}. The resulting radial chemical gradient reflects the average properties of stars at their radial birth locations, or \rbirth\ \citep{Minchev18,Frankel19,Buck20b}. However, since most stars have migrated from their original birth radii due to spiral arm and bar resonances \citep{SellwoodBinney,Roskar08,Minchev10,DiMatteo13}, we can expect a wide range of \rbirth\ values among stars observed today at any present-day galactocentric radii, \rcurrent. By tracing stars back to their \rbirth\ using chemistry and/or age based on models of the observed radial gradient \citep{Ness19,Lu22,Wang2024,Ratcliffe24}, we can more accurately assess how planet occurrence varies with a star’s birth environment. Without properly accounting for these relationships, there is a risk of conflating known correlations between stellar properties and planet occurrence (e.g., metallicity, age) with  the reality that those stellar properties independently reflect underlying Galactic chemical evolution.

\begin{figure*}
    \centering
    \includegraphics[width=\textwidth]{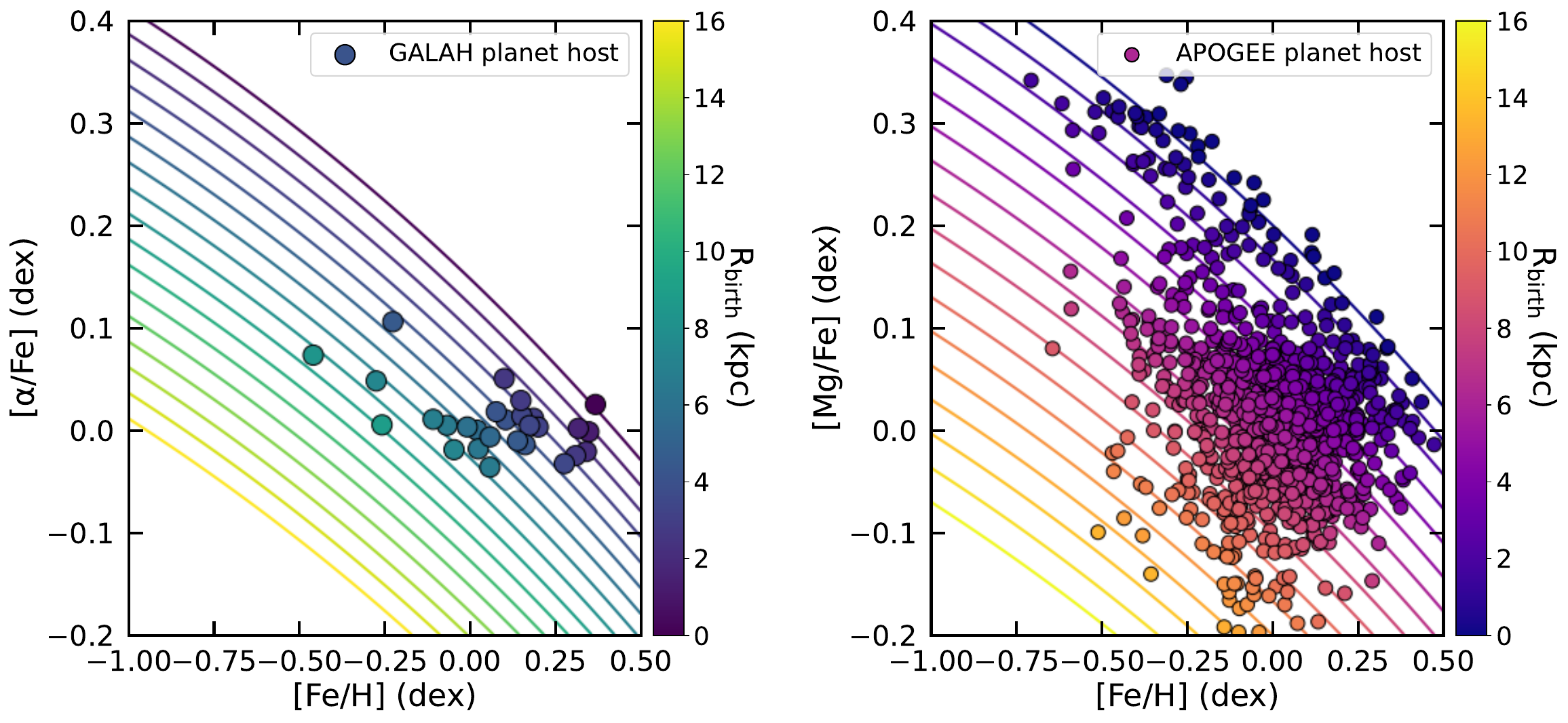}
    \caption{GALAH (left) and APOGEE (right) identified planet hosts in the \feh-\alphafe\ plane colored by inferred \rbirth. While most planet hosts were born at 5 kpc, just inward of the current solar neighborhood at 8 kpc, there are stars that have migrated from \rbirth\ of 0--13 kpc. The colored lines trace different chemical evolution sequences across the disk, as defined in Equations \ref{eqn:galahbr} and \ref{eqn:apogeebr} from \cite{Wang2024} and Mills et al. (in prep) with each line corresponding to a distinct \rbirth\ bin (in 2 kpc intervals). In this figure only, we show \rbirth\ tracks that are slightly shifted from those presented in Equations \ref{eqn:galahbr} and \ref{eqn:apogeebr} to reflect a more physically motivated \rbirth\ calibration. The rest of our plots and analysis use the unshifted calibration. We discuss the shift in Section \ref{sec:rbirth_recalc}, where we also show that the relative trends and core conclusions of our work are unaffected. }
    \label{fig:phs_rbirth}
\end{figure*}

In Figure \ref{fig:phs_rbirth}, we show calculated \rbirth\ tracks for Galactic Archaeology with HERMES (GALAH, \citealt{galah}) metallicity, \feh, and $\alpha$-enrichment, \alphafe, data on the left and Apache Point Observatory Galactic Evolution Experiment (APOGEE, \citealt{apogee}) \feh\ and \mg\ (a proxy for \alphafe) data on the right. Further described in Section \ref{sec:infer_rbirth}, these tracks were defined by \cite{Wang2024} and Mills et al. (in prep) and are inherited from the relationship seen in the \feh-\alphafe\ and \rbirth\ plane in cosmological simulations \citep{Buck20b}. While the rest of this work uses the original \rbirth\ calibration as defined in Equations \ref{eqn:galahbr} and \ref{eqn:apogeebr}, we adopt a slightly shifted version of these tracks in Figure \ref{fig:phs_rbirth} to reflect a more physically motivated \rbirth, as discussed in Section~\ref{sec:rbirth_recalc}. We note that shift does not affect the conclusions of this work, as the \rbirth\ calculation is inherently relative \citep{Lu22b}, and the relative trends we find remain unchanged regardless of the calibration chosen. We cross-match GALAH and APOGEE stars with the NASA Exoplanet Archive \citep{ps} using their Gaia IDs and show identified planet hosts from each sample (28 in GALAH, 1043 in APOGEE) and their position in the \feh-\alphafe\ plane, colored by their inferred \rbirth. Most stars appear to have been born just inward of the solar neighborhood (on average $5\pm 2$ kpc). However, there are stars that have migrated outward from 0 kpc and inward from 13 kpc, highlighting the wide radial Galactic coverage stars have migrated from to the solar neighborhood.


A planet host's presence in a particularly ``clustered" environment may also have dynamical or photoevaporation implications for its planets. In dense stellar environments, stars are more likely to experience dynamical encounters with nearby stars, such as stellar flybys, which can perturb planetary architectures (e.g., \citealt{Malmberg11,deJuan12,Cai18}). Additionally, stars in clustered environments may also be exposed to intense photoevaporation from nearby massive stars, which can strip material from protoplanetary disks suppressing planet formation (e.g., \citealt{Adams04,ConchaRamirez19}). \cite{Winter20} calculates the relative 5D and 6D phase-space (position and velocity) densities of planet hosts to assess if these stars are in such clustered environments and what impact this has on planet occurrence.

\subsection{Trends in Hot Jupiter Occurrence as a Test Case}
The inference of \rbirth\ relies on metallicity and $\alpha$-enrichment measurements \citep{Wang2024}, and phase-space density incorporates kinematic data \citep{Winter20}. Therefore, \rbirth\ and phase-space density are expected to manifest in planet occurrence patterns. Currently, planet occurrence is understood to correlate primarily with stellar metallicity. The relationship between stellar chemistry and kinematics provides a pathway to extend the planet frequency–metallicity correlation to Galactic scales. This framework can reveal which trends are truly driven by metallicity and which arise independently from \rbirth\ or phase-space density.


We use a series of toy models, grounded in well-established correlations between hot Jupiter occurrence and metallicity, to inject samples of stars with hot Jupiters and calculate their occurrence as a function of \rbirth\ and phase-space density. This allows us to explore how these correlations with metallicity manifest as a function of these Galactic-scale parameters. Since hot Jupiters show the clearest and strongest dependence on metallicity, they provide a key test case for disentangling known trends from Galactic-scale effects on planet occurrence. 

Hot Jupiters (mass range: $0.3 - 10\,M_{\mathrm{Jup}}$, orbital period range: 1-10 days) are a rare class of planets whose formation and migration mechanisms are still under debate \citep{Dawson18}, but they have been extensively studied due to detection biases. One of the most well-constrained findings is the planet frequency-metallicity relation \citep[e.g.,][]{Gonzalez97,Santos2004,Fischer05,Johnson10, Mortier13,Guo17,Adibekyan19,OsborneBayliss20}, where giant planets are more likely to be found around metal-rich stars. This supports the core accretion model of planet formation, where the cores of giant planets form more efficiently in metal-rich environments \citep[e.g.,][]{Pollack96}. 

Additionally, though beyond the scope of this work, recent studies have uncovered planet frequency-age correlations, showing that younger stars are more likely to host hot Jupiters, independent of metallicity correlations \citep{Chen23,MiyazakiMasuda23}. This is thought to be due to orbital decay effects over time leading to eventual planet engulfment (e.g., \citealt{Rasio96,Barker20}). Given hot Jupiters are found more frequently around younger stars and more metal-rich stars, this implies that hot Jupiters should predominantly be found in the Galaxy's low-$\alpha$ sequence. Low-$\alpha$ sequence stars are also inherently kinematically cooler \citep{Aumer09}. Thus, it is predicted and has been shown that hot Jupiters are found around younger, more metal-rich, and kinematically-cooler stars \citep{Adibekyan21,Mustill22,Blaylock24}.


\subsection{Toy Models as a Tool to Infer Planet Occurrence on Galactic Scales}
In this work, we present a framework to distinguish the planet frequency-metallicity correlation from larger-scale Galactic effects using a series of toy models. We inject hot Jupiters into a stellar sample based on known metallicity correlations, allowing us to explore how these trends propagate to hot Jupiter occurrence as a function of \rbirth\ and phase-space density. In Section \ref{sec:data}, we discuss the datasets from GALAH, APOGEE, and the NIHAO-UHD simulations \citep{Buck20b}, which we use to do our simulated planet injections. Section \ref{sec:methods} outlines how \rbirth\ and phase-space densities are inferred, how metallicity correlations are applied to the data as planet injections, and how planet occurrence is calculated. In Section \ref{sec:results}, we show the results of the toy model simulations and discuss their implications in Section \ref{sec:discussion}. We summarize our findings and conclude in Section \ref{sec:conclusion}.

\section{Data} \label{sec:data}
We explore hot Jupiter occurrence trends using the third data release from the GALAH survey \citep{galah, Wang2024} and the 17th data release from the APOGEE survey (\citealt{apogeedr17}). 
We compare these observational data to the simulated data from the NIHAO-UHD simulations \citep{Buck20b} that we treat as our ``complete" dataset. We limit our exploration to main-sequence and subgiant stars ($\logg > 3.44$), where hot Jupiter occurrence rates are most well-constrained. We also choose stars with chemistries consistent with the Galactic low and high-$\alpha$ sequence populations, where radial migration is applicable ($\alphafe > -0.2$, $\feh > -1$). Finally, we remove stars with large age errors ($> 50\%$) and non-physical inferred $\rbirth$ ($< 0$ or $>14$ kpc; beyond the calibration limits discussed in Section \ref{sec:infer_rbirth}).

We also limit the GALAH, APOGEE, and NIHAO-UHD stars to the same spatial coverage as the known planet hosts from the NASA Exoplanet Archive that were analyzed in \cite{Winter20}. Because these three datasets are not exoplanet surveys, they have much deeper survey coverage than the known exoplanet host population. This could result in different stellar populations compared to that from which the known exoplanet population is drawn, so we match the spatial distribution of these datasets to that of the known exoplanet population. Using the \texttt{SkyCoord} package in Astropy \citep{2013A&A...558A..33A,astropyii}, we calculate Galactocentric cartesian coordinates (X, Y, Z) for stars in GALAH, APOGEE, and the planet hosts listed in the NASA Exoplanet Archive based on their positions, parallaxes, proper motions, and radial velocities reported in the third data release (DR3) from Gaia \citep{GaiaDr3}. We use the available (X, Y, Z) for the NIHAO-UHD stars. We then select GALAH, APOGEE, and NIHAO-UHD simulation stars that fall within the X, Y, Z coordinate ranges of the planet hosts from \cite{Winter20} that are downselected from the NASA Exoplanet Archive based on their phase-space density classifications as discussed in Section \ref{sec:wintermethod}.

This selection effectively restricts each survey to the solar neighborhood, making them more analogous to the stellar samples from which exoplanet host stars are identified. These spatial cuts have varying impact on the fraction of stars in high/low-$\alpha$ sequence classifications (discussed below). With the cuts, 5\%, 13\%, and 30\% of stars belong to the high-$\alpha$ sequence in GALAH, APOGEE, and the NIHAO simulation, respectively. Compared to the parent samples, this results in no change for GALAH (5\%), a 2\% decrease for APOGEE (from 15\% to 13\%), and a 26\% decrease for NIHAO (from 56\% to 30\%). In the solar neighborhood, the thick disk population (corresponding to the high-$\alpha$ sequence) makes up $\sim 15-25\%$ of the stellar population \citep[e.g.,][]{Haywood19,Anguiano20}, so this brings the NIHAO dataset closer to expectations for the solar neighborhood. 

While the three datasets are never combined, we note the measurement offsets between APOGEE and GALAH are small. Based on the 465 stars in common between the two datasets, we find average differences of 0.02 dex in \feh, 7E-4 dex in \alphafe, and 0.94 kpc in calculated \rbirth. We now describe each dataset in more detail. 

\subsection{g2.79e12 NIHAO simulation}
The Milky-Way-like simulated data we use is from g2.79e12, a zoom-in simulation and part of the high resolution version of the NIHAO (Numerical Investigation of a Hundred Astronomical Objects) suite of cosmological hydrodynamical simulations designed to study the chemical, kinematic, and structural properties of Milky Way-mass galaxies \citep{Wang15, Buck20b}. This simulation represents a galaxy, consisting of 7.9 million stars, with a total virial mass of \(3.13 \times 10^{12} \, M_\odot\), a total stellar mass of \(15.9 \times 10^{10} \, M_\odot\), and a disk scale length of 5.57 kpc.

These simulations reproduce the \feh-\alphafe\ bimodality seen in the Milky Way, showcasing distinct high and low \alphafe\ sequences (using oxygen, [O/Fe], as the representative $\alpha$-process element) that correspond to the thick and thin disks respectively \citep[e.g.,][]{Buck20b,Buck21,Buck2023}. The high-$\alpha$ sequence forms early from rapid star formation and remains radially concentrated toward the galactic center with a large vertical scale height, while the younger, low-$\alpha$ sequence is more radially extended and confined to the disk mid-plane \citep[e.g.,][]{Buck2019}. Thus, it reproduces observed features of the Milky Way, such as metallicity gradients, age distributions, and relative scale heights of stellar populations. The outputs include stellar birth and present-day kinematics, metallicities, $\alpha$-abundances, and ages. 

Given the overall similarities in trends to the Milky Way \citep[e.g.,][]{Buck2018,Buck2019b}, this simulation (hereafter referred to as the NIHAO simulation and/or sample) serves as an important comparison and exploratory tool in this work, offering a statistically ``complete" dataset relative to the observed GALAH and APOGEE data. However, we do not treat it as an exact prescription or direct comparison for what to expect in occurrence space or in relation to the observed data, as it represents a Milky Way-\textit{like} galaxy. Notable differences include its higher total virial mass, larger stellar mass, and slightly extended disk scale lengths, which reflect a more dynamically heated and massive system compared to the Milky Way.

\begin{figure*}
    \centering
    \includegraphics[width=\linewidth]{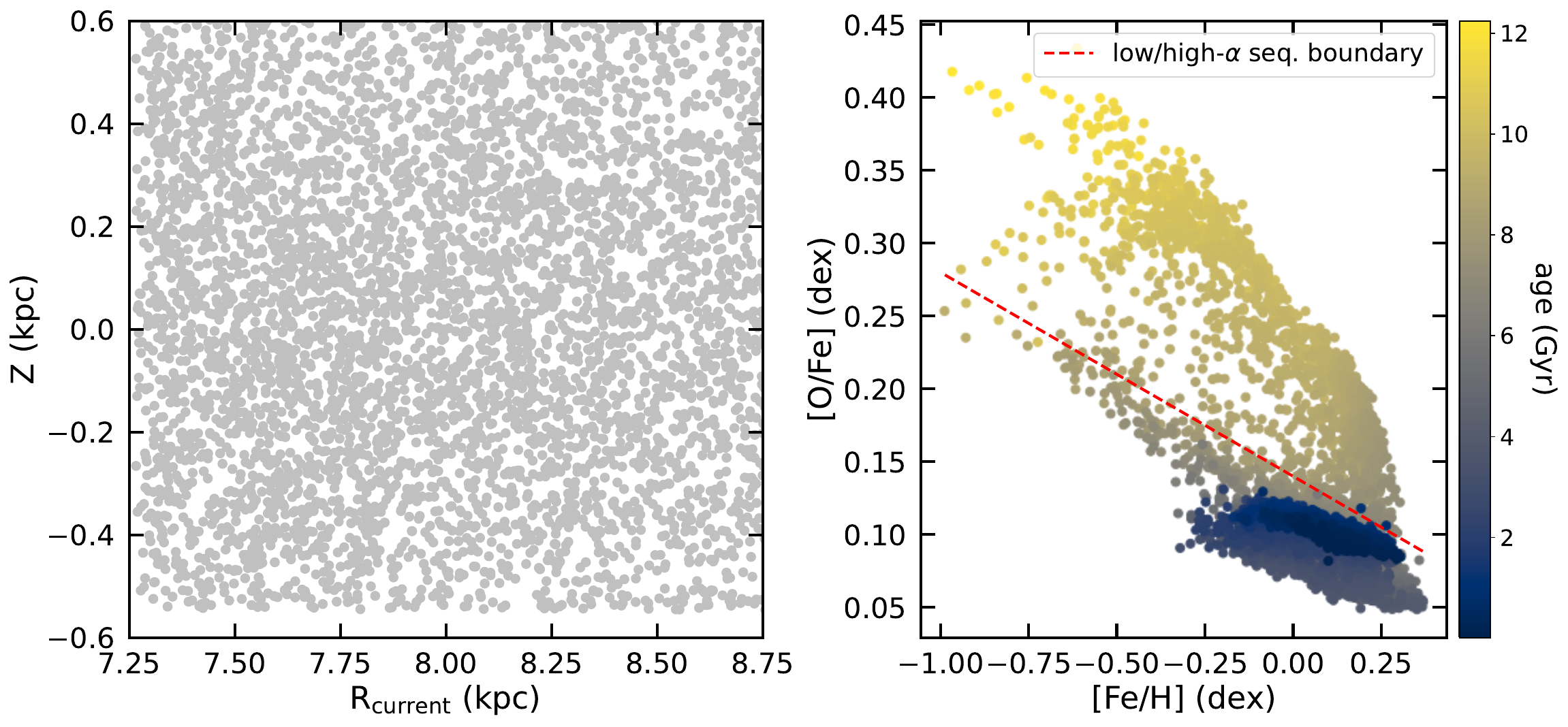}
    \caption{\rcurrent\ and Z coordinates (left) and \feh-[O/Fe] plane colored by their ages (right) of stars from the NIHAO-UHD \citep{Buck20b} simulation in the same spatial range as known planet hosts. We show the low and high-$\alpha$ sequence boundary with the red dashed line that was fit by eye following Figure 3 of \cite{Buck20b}}. The NIHAO sample is made up of 30\% high-$\alpha$ sequence stars. 
    \label{fig:b20dat}
\end{figure*}

After limiting the spatial coverage to match that of the known planet host population, we show the remaining 4,903 stars' positions in current galactocentric radius, \rcurrent, and Z space on the left and on the \feh-[O/Fe] plane colored by their ages on the right in Figure \ref{fig:b20dat}. Note the distribution of stars in \rcurrent-Z space is more ``complete" and evenly distributed compared to the distribution of stars in the GALAH and APOGEE samples shown in Figures \ref{fig:galah_hist} and \ref{fig:apogee_hist}. The low and high-$\alpha$ sequence populations are visually distinguishable. We separate the two disk populations with a line that we fit by eye following the demarcation shown in Figure 3 of \cite{Buck20b}. The low-$\alpha$ sequence population is below the line, and the high-$\alpha$ sequence population is above it. We draw similar lines by eye in the APOGEE and GALAH 
datasets to assign \textit{approximate} low and high-$\alpha$ sequence memberships. Of the 4,903 stars, 1,468 (30\%) are in the high-$\alpha$ sequence 

\subsection{GALAH}
The \cite{Wang2024} catalog uses data from the third data release from GALAH \citep{galah} to determine \rbirth (described in Section \ref{sec:infer_rbirth}). GALAH is a high-resolution spectroscopic survey (R = 28,000) that has observed 588,571 stars in its third data release across the Galaxy for galactic archaeology applications. This survey is conducted with the HERMES instrument on the Anglo-Australian Telescope \citep{galah_mission}. Stellar parameters are determined with 1D MARCS model atmospheres \citep{MARCSatmospheres} and the spectral synthesis code, Spectroscopy Made Easy \citep{SME1,SME2}. GALAH stars were cross-matched with those observed by the Gaia mission \citep{GaiaMission}, such that the value-added catalog includes stellar kinematics from the early third data release \citep{edr3} and stellar ages, which were inferred using isochrones from the Bayesian Stellar Parameter Estimation (BSTEP) code \citep{Sharma18}. \cite{Wang2024} applied the quality flag cuts recommended by the GALAH survey to create the highest fidelity sample. 

\begin{figure*}
    \centering
    \includegraphics[width=\linewidth]{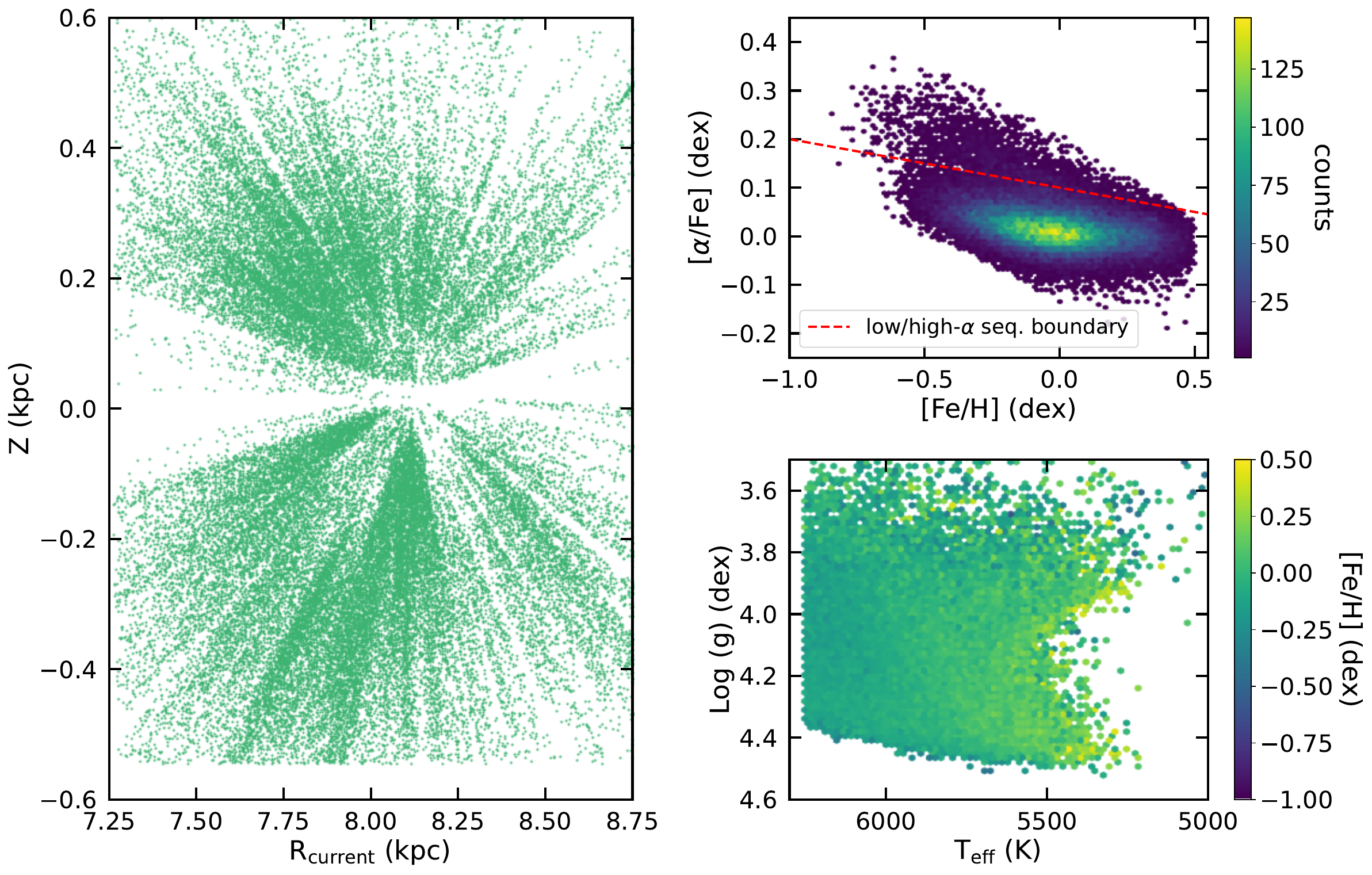}
    \caption{Overview of stellar properties in GALAH sample. \textit{Left:} \rcurrent\ and Z distribution. \textit{Top Right:} \feh-\alphafe\ plane colored by density with low and high-$\alpha$ sequence boundary shown by red line; 5\% of the GALAH sample's stars belong to the high-$\alpha$ sequence. \textit{Bottom Right:} Distribution of stars in the \teff-\logg\ plane colored by \feh.}
    \label{fig:galah_hist}
\end{figure*}

We update the kinematic measurements with those from Gaia's third data release \citep{GaiaDr3}. Gaia, launched in 2013, is a space observatory designed to create the most precise 3D catalog of stars in the Milky Way, providing astrometry over 1.46 billion stars. After making the cuts in \logg, chemical space, kinematics, and \rbirth\ described above, we are left with 49,427 stars. 
In Figure \ref{fig:galah_hist}, we show the stars' distribution in current \rcurrent-Z space (left), \teff-\logg\ plane colored by \feh\, (right, bottom), and \feh-\alphafe\ plane with the line separating the low and high-$\alpha$ sequences (right, top). From this approximate classification, 2,347/49,427 (5\%) stars are in the high-$\alpha$ sequence.  

\subsection{APOGEE}
The Mills et al. (in prep) catalog uses the 17th data release (DR17) of APOGEE \citep{apogeedr17} to infer \rbirth\ for 125,484 giant stars. APOGEE is another galactic archaeology-motivated high-resolution spectroscopic survey (R=22,500) in the near-infrared, primarily surveying red giants across the Galaxy \citep{apogee}. Stellar parameters for the $> 650,000$ stars in DR17 are derived using the APOGEE Stellar Parameter and Chemical Abundances Pipeline (ASPCAP) pipeline that fits spectra using synthetic stellar atmosphere models \citep{aspcap}.

\begin{figure*}
    \centering
    \includegraphics[width=\linewidth]{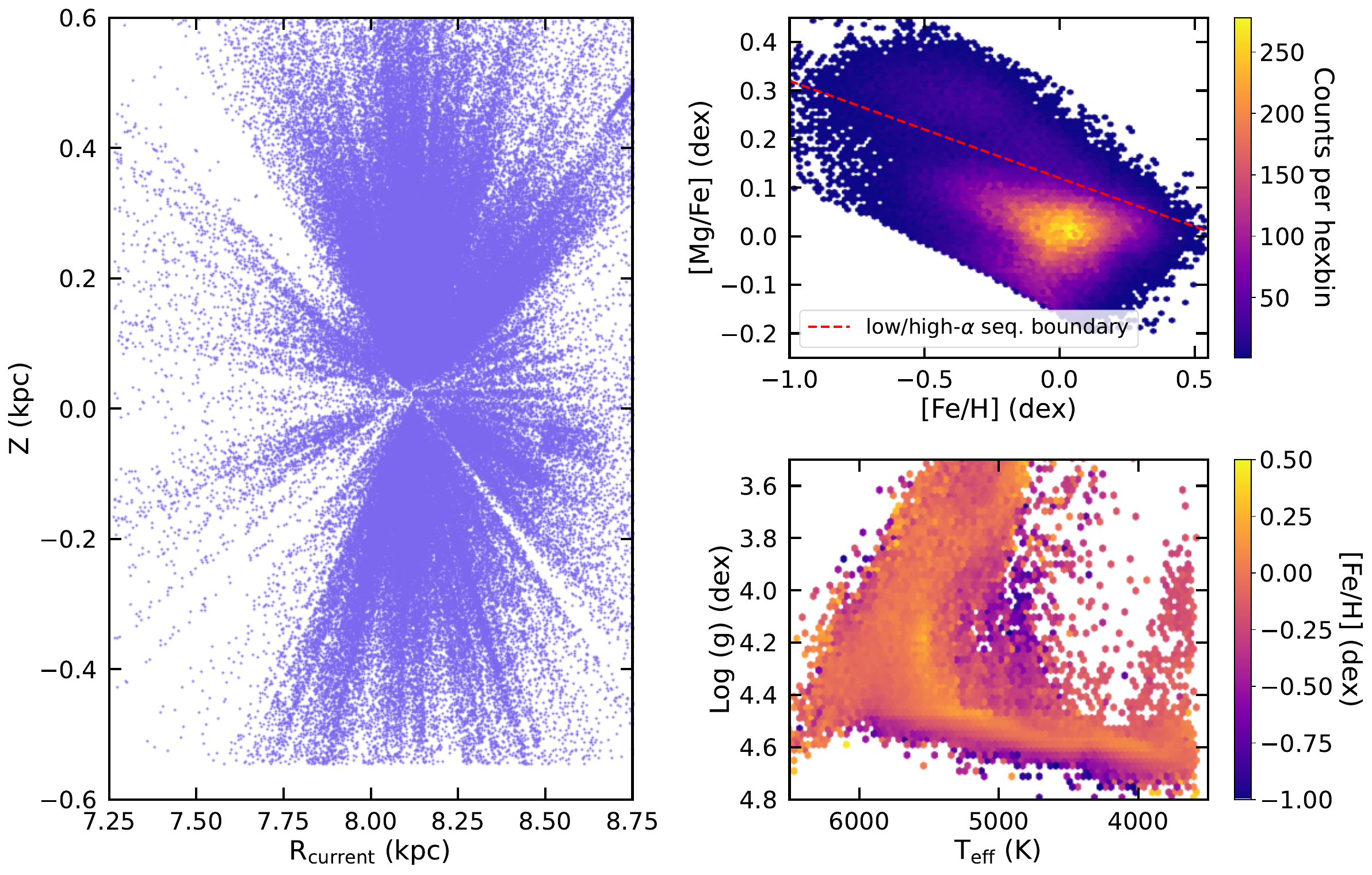}
    \caption{Overview of stellar properties in APOGEE sample. 
    \textit{Left:} \rcurrent\ and Z distribution. \textit{Top Right:} \feh-\mg\ plane colored by density with low and high-$\alpha$ sequence boundary shown by red line. 13\% of the APOGEE sample's stars belong to the high-$\alpha$ sequence. \textit{Bottom Right:} Distribution of stars in the \teff-\logg\ plane colored by \feh.}
    \label{fig:apogee_hist}
\end{figure*}

We calculate \rbirth\ for 252,441 dwarf and subgiant stars from APOGEE DR17 using [Fe/H]-[Mg/Fe], following the Mills et al calibration. This calculation is outlined in Section \ref{sec:infer_rbirth} and Equation \ref{eqn:apogeebr}. We cross-match these stars in Gaia DR3, so we have the stars' kinematics \citep{GaiaDr3}. We also cross-match these stars with one of APOGEE's value-added catalogs, ASTRO-NN \citep{astroNN}, for inferred stellar ages. ASTRO-NN uses deep learning methods trained on bright asteroseismic targets in the data to infer stellar parameters, including stellar ages \citep{apogee_ages}. After removing stars with ``bad star" flags we are left with 138,255 stars. In Figure \ref{fig:apogee_hist}, we show the stars' distribution in current \rcurrent-Z space (left), \teff-\logg\ plane colored by \feh\ (right, bottom), and \feh-\mg\ plane with the line separating the low and high-$\alpha$ sequences (right, top). From this approximate classification, 17,989/138,255 (13\%) stars are in the high-$\alpha$ sequence.

\section{Methods} \label{sec:methods}
\subsection{Determining \rbirth} \label{sec:infer_rbirth}
To understand how metallicity correlations manifest in planet occurrence as a function of \rbirth, we need to infer \rbirth\ for our sample of stars. We outline how \rbirth\ is inferred for the APOGEE and GALAH catalogs from \cite{Wang2024} and Mills et al. 2024 (in prep). \cite{Wang2024} employ an empirical method of inferring \rbirth\ using \feh\ and \alphafe\ measurements for $\sim 59,000$ GALAH stars informed by trends in cosmological hydrodynamic simulations from the NIHAO-UHD project \citep{Buck20}. These zoom-in simulations of g2.79e12, a Milky Way type galaxy, show trends with \rbirth\ in the \feh-[O/Fe] plane. The gradients are so clear that individual \rbirth\ ``tracks" can be placed down (see Figure 3 of \citealt{Buck20b}). This reveals that stars higher in [O/Fe] come from the inner regions of the Galaxy, and those that have lower [O/Fe] ratios are spread out over larger \rbirth. Since [O/Fe] is an $\alpha$-process element commonly used as a proxy for \alphafe, and \alphafe\ itself is a proxy for stellar age, with $\alpha$-enriched stars typically being older, these trends reinforce the Galaxy's chemical evolution. Older stars formed closer to the Galactic center. Motivated by the \rbirth-abundance ratio trends, \cite{Wang2024} lay down the same \rbirth\ tracks from 0 to 14 kpc on GALAH data in the \feh-\alphafe\ plane switching to \alphafe\ since it is better measured in GALAH data. These \rbirth\ tracks were fit from the simulation data and are empirically described with the following equation: 
\begin{equation} \label{eqn:galahbr}
\begin{aligned}
R_{\text{birth,GALAH}} =\ & -40 \times ([\alpha/\text{Fe}] + 0.80 \\ & \times \exp(0.4 \times [\text{Fe}/\text{H}]) - 0.81) + 8.
\end{aligned}
\end{equation}

Mills et. al (in prep) apply the same approach to $\sim 125,000$ APOGEE stars\footnote{Although this calibration was developed for APOGEE giants, applying it to main sequence and subgiant stars is valid since \rbirth\ depends on orbital dynamics not internal stellar structure \citep{SellwoodBinney}.} and infer \rbirth\ with the following equation: 
\begin{equation} \label{eqn:apogeebr}
\begin{aligned}
R_{\text{birth,APOGEE}} =\ & -30 \times ([\alpha/\text{Fe}] + 0.80 \\ & \times \exp(0.4 \times [\text{Fe}/\text{H}]) - 0.81) \\ & + 8.89,
\end{aligned}
\end{equation}
where \mg\ is used as a representative \alphafe\ element.

We use the reported \rbirth\ from the NIHAO simulation and apply Equations \ref{eqn:galahbr} and \ref{eqn:apogeebr} to calculate \rbirth\ for our GALAH and APOGEE samples. We calculate \rbirth\ for each star using Monte Carlo sampling. We sample from two Gaussian distributions defined by the stars' \feh\ and \alphafe\ and their reported uncertainties 1000 times to calculate the resulting \rbirth\ distribution. We adopt the mean and standard deviation of this distribution as the star's \rbirth\ value and its associated uncertainty.

\subsection{Calculating Phase-Space Densities of Stars}\label{sec:wintermethod}
We are also interested in seeing metallicity correlations manifest in planet occurrence as a function of phase-space density. Here, we replicate the exact methods described in \cite{Winter20}, which we summarize here. The authors calculate the probability of planet hosts being in relative phase-space densities within the Galaxy. For each planet host, they find all the stars within a 40 pc radius in Gaia DR2 \citep{gaiadr2} and randomly select between 400--600 of these stars  and their associated kinematics (RA, Dec, parallax, proper motions, and radial velocities if available). They then calculate the Mahalanobis distance, a quantity that describes the separation between an exoplanet host star and its neighbors in the multidimensional phase-space. The phase-space density is calculated as the inverse of this distance to the star's 20th nearest neighbor and normalized by the median density of all the stars in the neighborhood. This provides a relative measure of local stellar clustering. They then apply a Gaussian mixture model to these relative phase-space densities to identify distinct overdense and underdense regions of phase-space. The outputs from the model are the planet hosts' probabilities of membership in high-density/overdense or low-density/underdense environments. The authors require a probability of $>84\%$ in either type of density for stars to have a clear classification that can be used in their analyses. 

\begin{figure*}
    \centering
    \includegraphics[width=\linewidth]{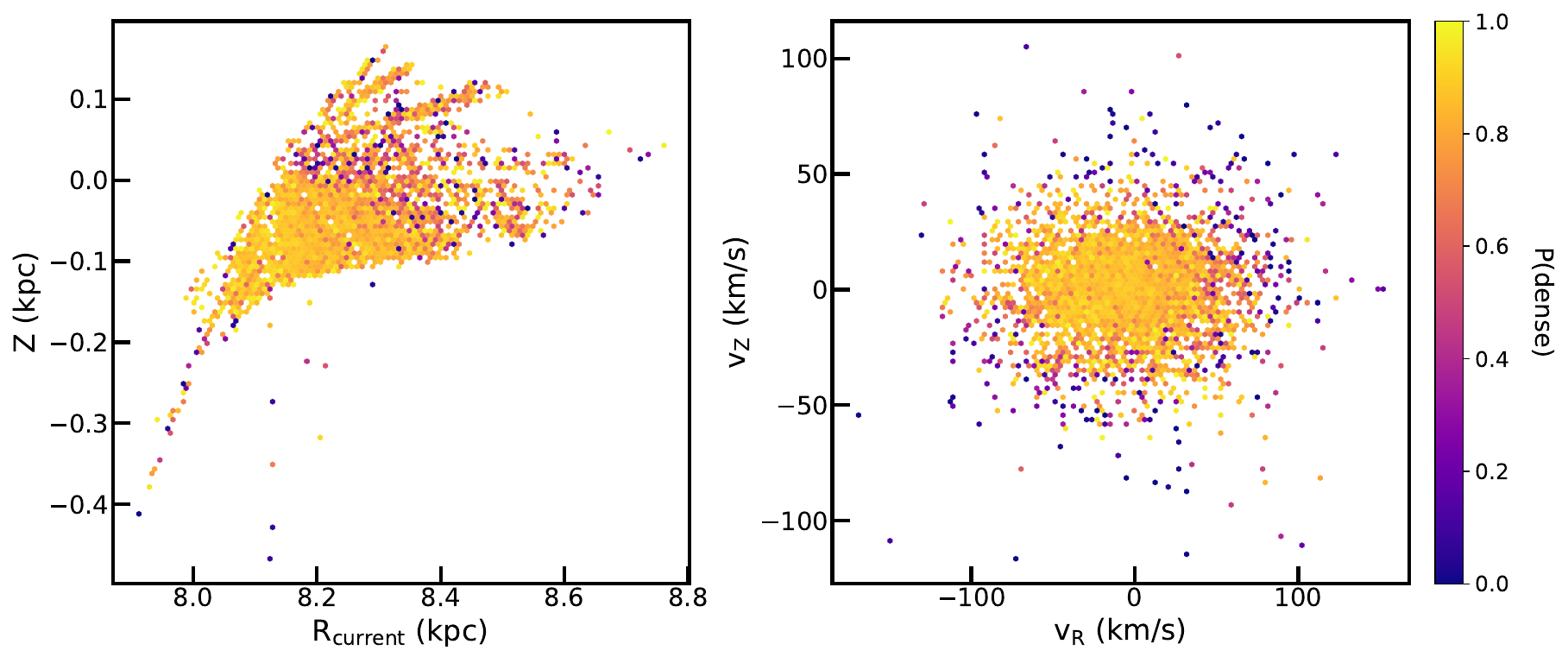}
    \caption{Spatial and kinematic distribution of APOGEE stars colored by calculated phase-space densities following calculation and visualizaiton outlined in Figure 1 of \cite{Winter20}. \textit{Left:} R$_{\rm current}$-Z spatial distribution  APOGEE catalog colored by P(dense). \textit{Right:} V$_{\rm R}$-V$_{\rm Z}$ velocity distribution colored by P(dense). The stars classified to be in the overdense environments are evident given their close proximity to neighbors spatially and kinematically compared to the stars classified to be in underdense environments.}
    \label{fig:winter_vis}
\end{figure*}

We follow the method \cite{Winter20} outlined above to calculate the probability of high density and low density phase-space membership for our GALAH, APOGEE, and NIHAO stars using the authors' publicly available code\footnote{Their code can be found in the following Github repository: \href{https://github.com/ajw278/astrophasesplit}{https://github.com/ajw278/astrophasesplit}.} and Gaia DR3 \citep{GaiaDr3}. For each NIHAO star in the simulation, we do not query Gaia DR3 as we do with the other datasets. Instead, we simply randomly choose 400-600 stars within a 40 pc radius from the simulated dataset itself. This approach ensures consistency within the simulation framework, as the simulated Galaxy has inherent differences from the actual Milky Way. In Figure \ref{fig:winter_vis}, we show the APOGEE stars spatial and kinematic distributions colored by their calculated phase-space densities (or P(dense)). It is evident that stars located in crowded spatial and kinematic regions yield higher phase-space densities. A visualization of the calculation itself can be found in Figure 1 of \cite{Winter20}.

\subsection{Applying Hot-Jupiter Metallicity Power Laws}
Given hot Jupiter occurrence trends with metallicity and age, \cite{Chen23} fit power-laws to describe these relations that we use in this work. The metallicity and kinematic information in their analysis comes from the first Planets Across Space and Time (PAST) catalog \citep{PASTI}, which includes a combination of APOGEE, Gaia, Large Sky Area Multi-Object Fiber Spectroscopic Telescope (LAMOST, \citealt{LAMOST}), and RAdial Velocity Experiment (RAVE, \citealt{ravedr5}) measurements. The \cite{Chen23} study consisted of 355 stars with 193 hot Jupiters identified from the NASA Exoplanet Archive. They separate these planet samples into the different detection methods used to characterize each system: space-based transits, ground-based transits, and radial velocities. 

The authors constructed three power law models to model the relationship between planet occurrence and metallicity, age, and the combination of these two parameters. A Bayesian analysis was employed to vary the model parameters and evaluate the likelihood functions by comparing the predicted occurrence rates from these power laws to the observed data. Best-fit parameters were determined by maximizing the likelihood, and the posterior distributions of these parameters were used to estimate uncertainties. The best-fit parameters given each type of detection method can be found in their Table S4. We make use of their metallicity power law in this work. 

Their metallicity-dependent power law, $F_{\text{HJ}}(\feh)$, is represented as follows: 
 \begin{equation} \label{eqn:fehprop}
     F_{\text{HJ}}(\feh) \propto 10^{\beta \cdot [\text{Fe/H}]},
 \end{equation}
 where $\beta$ describes the strength of the dependence of \feh. 
 


While the authors provide average best-fit parameters for the metallicity-only power law across detection method samples, they do not explicitly calculate a normalization constant, $C$ for these models. Thus, for the metallicity-only power law, we assume a normalization constant of 0.01 since on average, solar-metallicity stars have a 1\% hot Jupiter occurrence rate (\citealt{Beleznay22} and references therein). Expression \ref{eqn:fehprop} becomes 
\begin{equation}\label{eqn:fehPL}
    F_{\text{HJ}}(\feh) = 0.01 \times 10^{1.6 \cdot [\text{Fe/H}]},
\end{equation}

 with the reported $\beta = 1.6$ from \cite{Chen23}. 
 
 Using Equation \ref{eqn:fehPL}, we calculate the planet occurrence probability for each star in the GALAH, APOGEE, and simulated NIHAO datasets based on their metallicity. 
 We calculate and compare occurrence rates using power laws with varying $\beta$ from (0.6--3.4) outlined in Table 1 of \cite{OsborneBayliss20} and references therein. We find that while the absolute occurrence rates can vary by $-3-+86\%$, the overall trends explored in this work remain qualitatively unchanged. Because our goal is to examine how occurrence varies across Galactic parameters in a toy model framework, a precise normalization of $\beta$ is not strictly required.
 
The steep metallicity dependence of the power law naturally reproduces the theorized suppression of giant planet formation at low \feh\ $\lesssim -0.7$ dex \citep{Johnson12,Hasegwa14}. While we do not impose a hard threshold, our selection of stars with \feh\ $> –1$ to avoid halo stars ensures that relatively few metal-poor stars are included in the sample. Across 10,000 planet injection trials, we find that on average only 0.02\% of injected planets orbit stars with $\feh < –0.7$. We also note that the FGK stars and M dwarfs in APOGEE probe similar Galactic populations, so we do not need to consider the dependencies of hot Jupiter occurrence on spectral type \citep[e.g.,][]{Johnson10,Maldonado20,Sabotta21,Beleznay22}.


\subsection{Planet Injection \& Planet Occurrence Calculations}
 We probablistically inject planets into the sample by generating a random number between 0 and 1 for each star. If the randomly drawn number is less than the calculated probability from Equation \ref{eqn:fehPL}, a planet is ``injected", and the star is tagged as a planet host. To calculate occurrence rates as a function of \rbirth\ we count how many stars are planet hosts versus not in \rbirth\ bins of 1 kpc. When considering relative hot Jupiter occurrence rates across phase-space densities, we compare how many hot Jupiter planet hosts are found in phase-space under-densities and over-densities relative to the total population, as was done in \cite{Winter20}.

\section{Results} \label{sec:results}
\subsection{Planet Occurrence as a Function of \rbirth}\label{sec:rbirthrsults}

\begin{figure*}
    \centering
    \includegraphics[width=\textwidth]{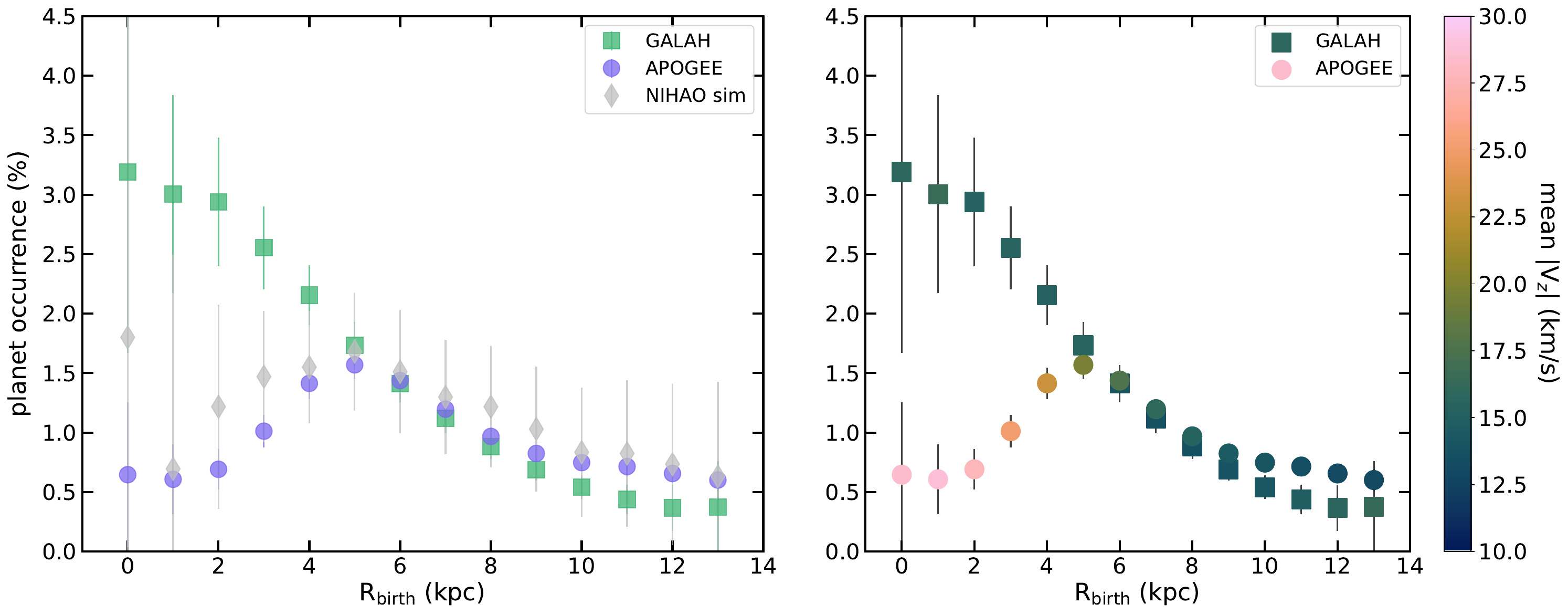}
    \includegraphics[width=\textwidth]{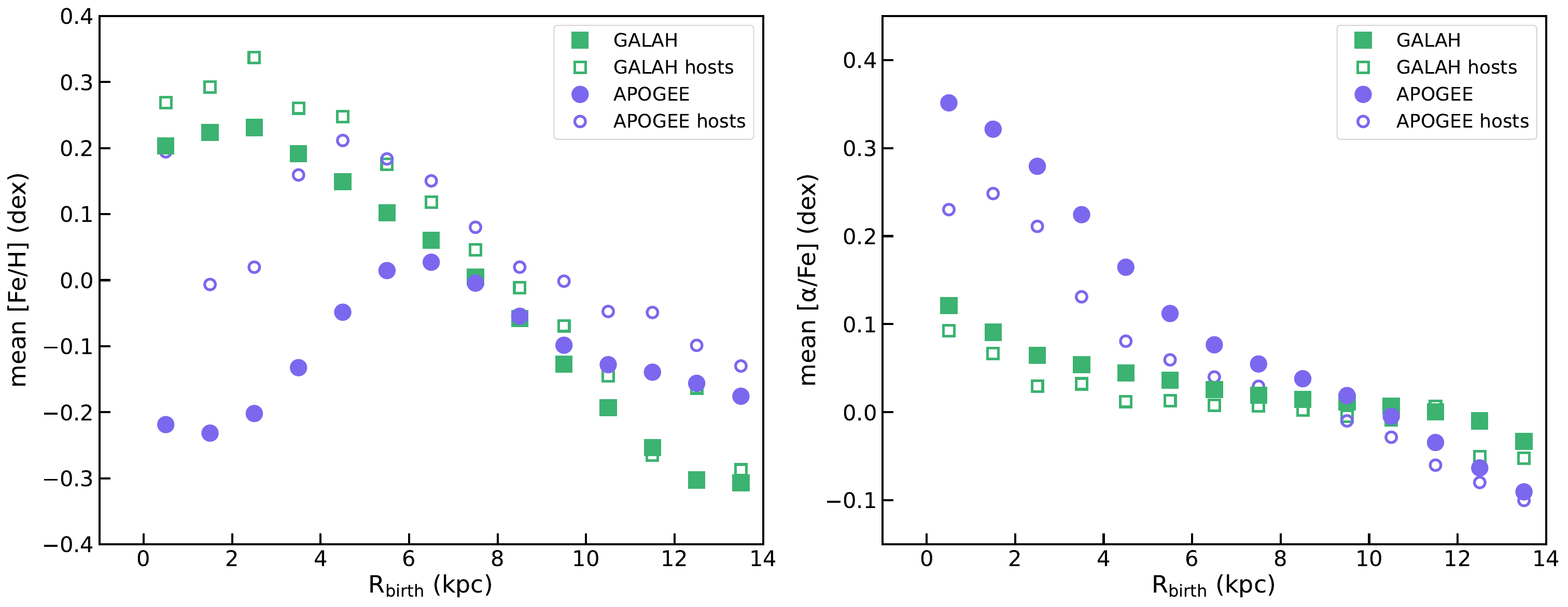}
    \caption{Simulated hot Jupiter occurrence as a function of \rbirth\ based on the hot Jupiter-metallicity power law from \cite{Chen23}. Mean occurrence and standard deviation of 1000 sets of injections are plotted. \textit{Top Left:} GALAH (green squares), APOGEE (purple circles), and NIHAO simulation (gray diamonds) data show decreased rates of hot Jupiter occurrence from 5--14 kpc. The NIHAO simulation and APOGEE distributions are similar. \textit{Top Right:} GALAH and APOGEE data colored by mean absolute vertical velocity. Discrepancies in planet occurrence between the two surveys at \rbirth\ $< 5$ kpc are correlated to differences $> 5$ km/s in mean stellar vertical velocities, where APOGEE stars are kinematically hotter. \textit{Bottom Left:} Mean \feh\ binned by \rbirth\ for GALAH and APOGEE sample as solid markers and planet host population as unfilled markers. The mean metallicity reflects the simulated planet occurrence trends in the top left panel, where APOGEE stars with \rbirth $< 5$ kpc are more metal-poor by 0.2--0.4 dex compared to GALAH stars at the same \rbirth. \textit{Bottom Right:} Mean \alphafe\ binned by \rbirth\ for GALAH and APOGEE sample as solid markers and planet host population as unfilled markers. APOGEE stars with \rbirth $< 5$ kpc are more $\alpha$-enriched by 0.2 dex compared to GALAH stars at the same \rbirth. The net increase in mean \feh\ in the planet host population versus the entire sample reflects the planet frequency -metallicity power law used to inject stars with hot Jupiters. Because of this exponential weighting, even small shifts in the distribution of metal-rich stars in APOGEE across \rbirth\ shift the peak of planet occurrence and thus the mean \feh\ of the planet host population inward by 2 kpc independently of the peak in mean \feh\ at 7 kpc.} 
    \label{fig:feh_inj}
\end{figure*}

We simulate hot Jupiter occurrence as a function of \rbirth. Figure \ref{fig:feh_inj} shows the mean hot Jupiter occurrence and standard deviation from 1000 sets of injections. In the top left panel, we compare data from the GALAH, APOGEE, and the NIHAO simulation. Hot Jupiter occurrence increases towards the inner regions of the Galactic disk (until 5 kpc). 
All three samples are in agreement $\geq 5$ kpc. APOGEE (shown in purple circles) and the NIHAO simulation data (shown in gray diamonds) are in agreement across all \rbirth.

The discrepancies between GALAH (shown in green circles) and APOGEE planet occurrence rates at \rbirth $< 5$ kpc likely arise due to differences in GALAH and APOGEE's target selection. We show the mean \feh\ and \alphafe\ for each \rbirth\ bin in the bottom row of Figure~\ref{fig:feh_inj}, for the entire GALAH and APOGEE samples as the filled green squares and purple circles and the planet hosts as the unfilled markers. The net increase in mean \feh\ for the planet host samples reflects the injection of planets into stars based on the planet frequency-metallicity power law from Equation \ref{eqn:fehPL}. For APOGEE, this exponential dependence of planet frequency on \feh\ shifts the peak in occurrence and the mean \feh\ of the planet host population inward from the peak in the full sample’s mean \feh. The full sample peaks at 7 kpc, while the planet occurrence and planet host mean \feh\ both peak around 5 kpc. Although both trends are shaped by the underlying \feh\ distribution, the peak in planet occurrence is shifted inward from the peak in mean metallicity due to the exponential dependence of occurrence on \feh, which gives greater weight to the highest-metallicity stars even if they are relatively fewer in number at that \rbirth\ bin. At \rbirth $< 5$ kpc, GALAH stars are on average $> 0.2$ dex more metal-rich and $\sim 0.2$ dex more depleted in \alphafe\ than APOGEE stars. The lower mean \feh\ for APOGEE stars at \rbirth $< 5$ kpc sets the lower occurrence rates compared to GALAH. These mean differences in \feh\ correlate with other variables, including kinematics. 
As shown in the top, right panel, the difference in planet occurrence appears when the mean absolute vertical velocities, V$_{\text{Z}}$, between the two samples exceeds differences of 5 km/s. At \rbirth\ $< 5$ kpc, the GALAH stars are kinematically cooler than APOGEE and NIHAO simulation stars. 

We note that although the NIHAO data matches APOGEE in planet occurrence as a function of \rbirth, we do not show its velocity and chemistry distributions in the other panels. Given the NIHAO simulation is a Milky Way-\textit{like} simulation, these distributions quantitatively differ in ways that detract from the main results and are thus not helpful to visualize.

\subsection{Planet Occurrence as a Function of Phase-Space Density}

\begin{figure*}
    \centering
    \includegraphics[width=\linewidth]{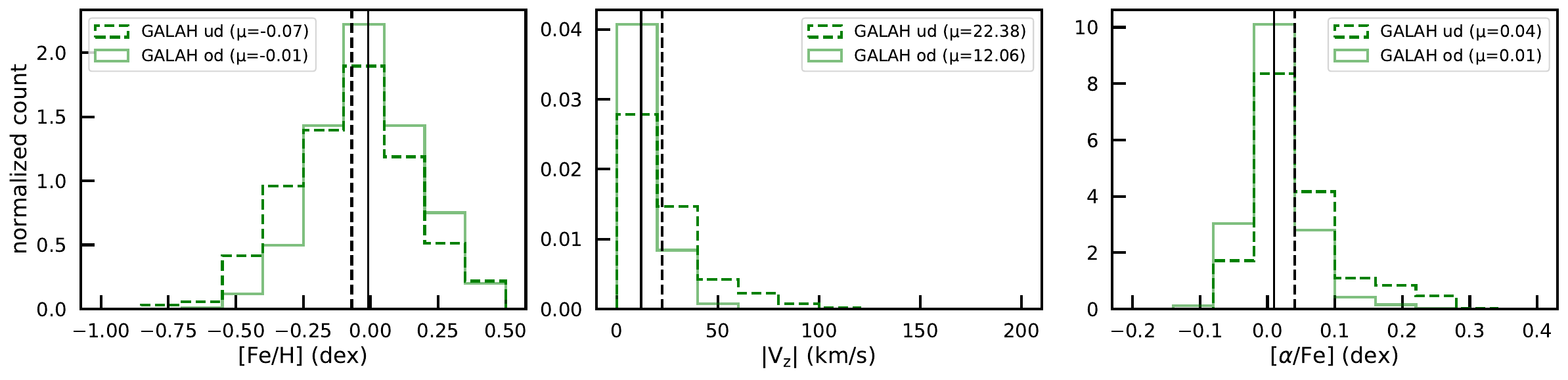}
    \includegraphics[width=\linewidth]{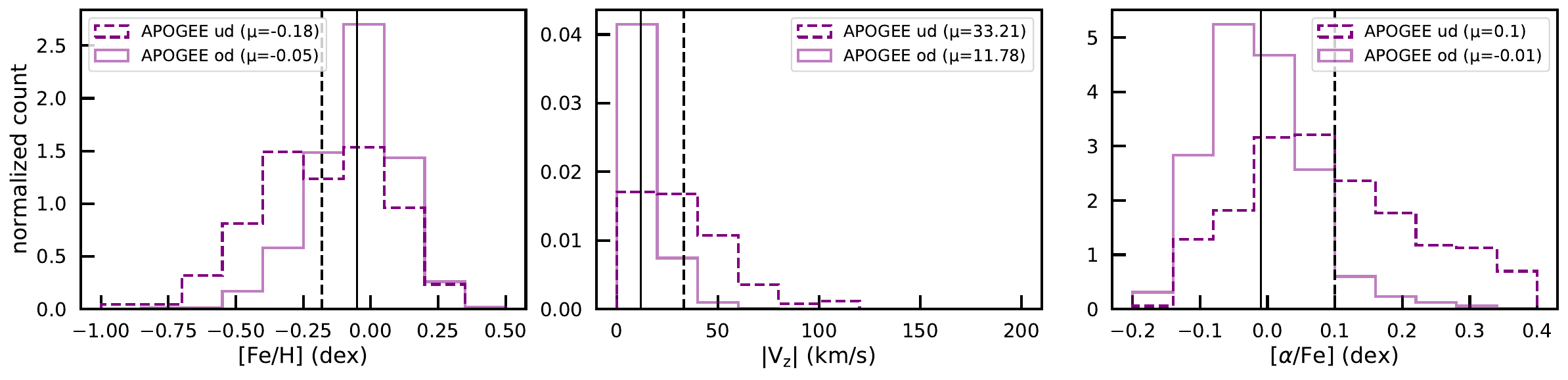}
    \includegraphics[width=\linewidth]{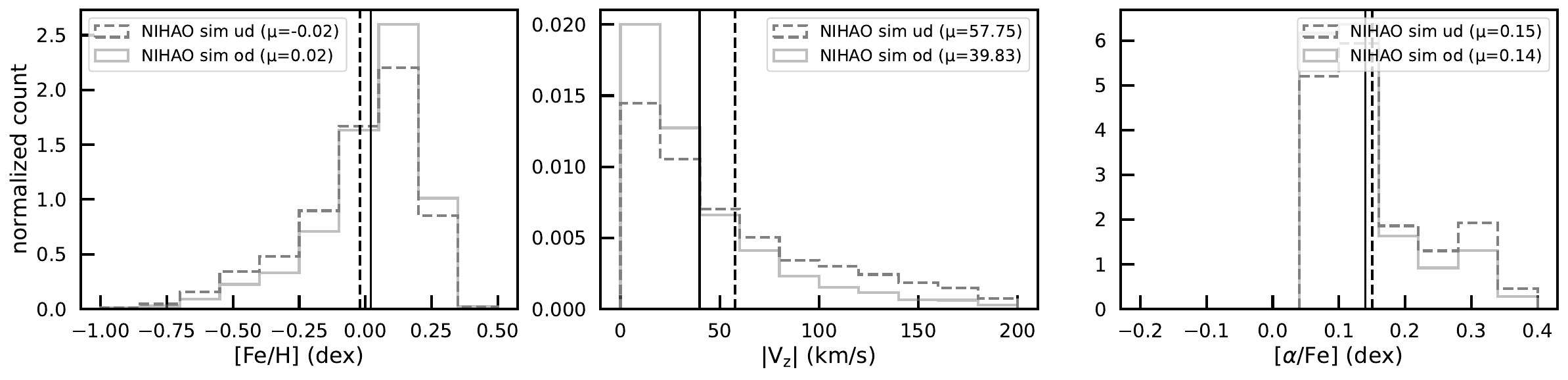}
    \caption{Distributions of GALAH (in green), APOGEE (in purple), and NIHAO simulation (in gray) stars in overdense environments (shown as a solid line histogram and abbreviated as od) to those in underdense environments (shown as a dashed line histogram and abbreviated as ud) in \feh-, absolute V$_{\rm Z}$-, and \alphafe-space as classified by the analysis done in \cite{Winter20}. The means of each distribution of stars in overdensities and underdensities are denoted by solid and dashed vertical black lines, respectively. These distributions are for all stars, regardless of planet host status. For this parent population, stars in underdensities are on average more metal-poor by 0.04--0.13 dex, kinematically-hot by 10--22 km/s, and $\alpha$-enriched by 0.01--0.1 dex compared to stars in overdensities. }
    \label{fig:parentpop}
\end{figure*}

\begin{figure*}
    \centering
    \includegraphics[width=0.88\textwidth]{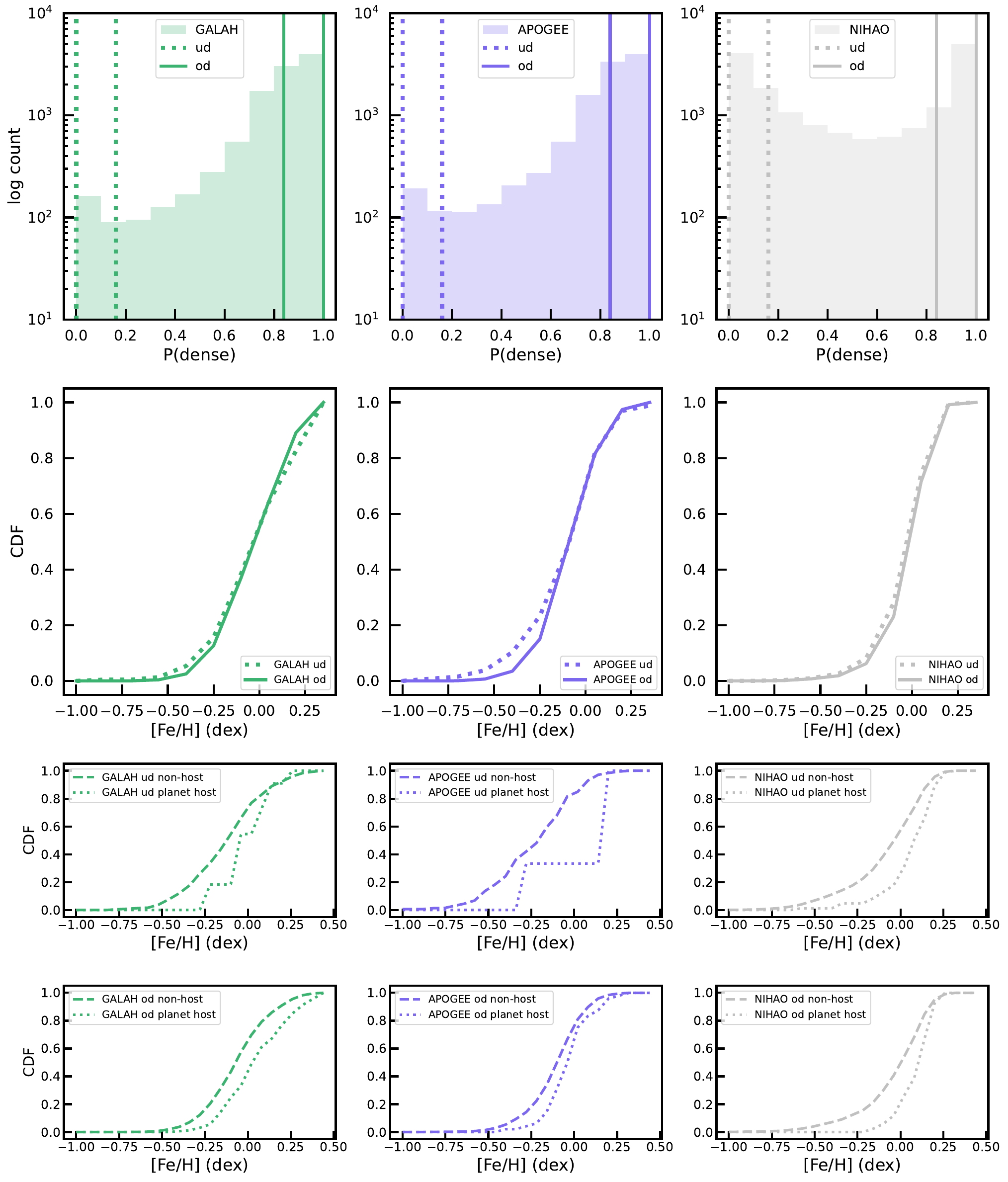}
    \caption{\cite{Winter20} phase-space density analysis with injected hot Jupiters based on the hot Jupiter occurrence-metallicity power law from \cite{Chen23}. \textit{Top:} Histograms of simulated hot Jupiters in phase-space underdensities (dotted lines and calculated P(dense) $< 0.16$) and overdensities (solid lines and calculated P(dense) $> 0.84$) shown for GALAH (green, left), APOGEE (purple, middle), and the NIHAO simulation (gray, right).  We are able to replicate \cite{Winter20}'s result, where $>$ 95\% of hot Jupiter hosts are in overdensities compared to underdensities for GALAH (\god\ versus \gud) and APOGEE (\aod\ versus \aud). The distribution of hot Jupiters in the NIHAO simulation is more evenly split (\nod\ versus \nud) for reasons we discuss in Section \ref{sec:simdifs}. \textit{Middle:} Cumulative distribution functions (CDFs) in underdensities (dotted lines) and overdensities (solid lines) following Figure 3 of \cite{Winter20} in \feh. \textit{Bottom:} \feh\ distributions of stars with and without injected hot Jupiters for each underdense (top) and overdense (bottom) population per dataset. The planet host CDFs for under/overdensities are very similar in \feh-space with average mean differences $\leq 0.02$ dex compared to the planet and non-planet host differences for each environment (0.04--0.12 dex) as shown in the two bottom panels here and in the left column of Figure \ref{fig:parentpop}.}
    \label{fig:winter}
\end{figure*}

Following the analysis done in \cite{Winter20} and described in Section \ref{sec:wintermethod}, we calculate the phase-space densities for GALAH, APOGEE, and NIHAO simulation stars. We find of the 49,427, 138,255, and 4,903 stars, 15,025 (30\%), 59,450 (43\%), and 4,903 (100\%) stars 
are able to be classified in that there are enough stars within a 40 pc radius to calculate relative phase-space densities. Of those classifiable stars, we find 8,559 (59\%), 27,941 (47\%), and 1,699 (35\%) stars in overdensities and 373 (2\%), 3,427 (6\%), and 1,616 stars (33\%) in underdensities for each sample. Recall that \cite{Winter20} defined stars with bona-fide overdensity or underdensity classifications as those with calculated densities, or P(dense), $>$ 84\% or $< $ 16\%. We repeat the density calculation at least $10$ times to build up a robust dataset to explore. Each time a star's phase-space density is calculated, a different set of random, neighboring stars will be selected, so a star's calculated phase-space density may vary. We find stars' phase-space densities vary on average 4\% for GALAH and APOGEE and 72\% for NIHAO. The high variability in phase-density calculations for NIHAO is explained in Section \ref{sec:simdifs}.

We compare the distributions of all GALAH (in green), APOGEE (in purple), and NIHAO simulation (in gray) stars in overdense environments (shown as a solid line histogram) to those in underdense environments (shown as a dashed line histogram) in \feh-, absolute V$_{\rm Z}$-, and \alphafe-space in Figure \ref{fig:parentpop}, regardless of planet host status. In all 3 samples, we see separations in the overdensity and underdensity populations. For GALAH, APOGEE, and NIHAO: stars in underdensities are on average more metal-poor by 0.06, 0.13, and 0.04 dex, kinematically-hot by 10, 22, and 18 km/s, and $\alpha$-enriched by 0.03, 0.1, and 0.01 dex compared to stars in overdensities. We assess the statistical robustness of the \feh\ and \alphafe\ separations in GALAH and APOGEE by bootstrap resampling using the measurement errors and confirm that the distributions remain visually distinct.

We then test metallicity correlations with hot Jupiter occurrence as a function of phase-space over/underdensities. \cite{Winter20} show that 92\% of hot Jupiters are found around stars in phase-space overdensities. To mirror their methodology, we inject hot Jupiters into the GALAH, APOGEE, and NIHAO samples and repeat this injection process 20 times to build up a robust dataset. With each injection, we calculate phase-space densities for the entire sample and filter for planet-hosting stars with confident classifications (P(dense) $> 0.84$ or $< 0.16$), as done in \citet{Winter20}. We then compute the fraction of planet hosts in overdensities versus underdensities out of the total number of classified injected planet hosts. After calculating phase-space densities for $> 20$ sets of injected hot Jupiter hosts, we find that $>$95\% are in overdensities (solid lines) around \god\ GALAH (in green, left panel) and \aod\ APOGEE (in purple, middle panel) stars compared to the \gud\ GALAH and \aud\ APOGEE stars in underdensities (dotted lines) as shown in Figure \ref{fig:winter}. The NIHAO injected planets (in gray, right panel) are more evenly split between stars found in overdensities (52\%; \nod\ planet hosts) and underdensities (48\%; \nud\ planet hosts) differing from the GALAH and APOGEE results for reasons we explain in Section \ref{sec:simdifs}. 

We also show the cumulative distribution functions (CDFs) for the injected hot Jupiters in underdensities and overdensities in \feh-space in the middle row of Figure \ref{fig:winter}.  
We see similar CDF shapes for both overdensities and underdensities. While the hot Jupiter hosts in underdensities do visually appear on average slightly more metal-poor, the \feh\ distributions of the planet hosts in overdensities and underdensities are much more similar ($\leq 0.02$ dex) than what we see the overall samples shown in Figure \ref{fig:parentpop} (0.04--0.13 dex). This is visually validated in the bottom two panels of Figure \ref{fig:winter}, where we show the planet host and non planet-host populations for each phase-space density environment. We see that the planet host population is always higher in \feh\ compared to the non planet-host population, which is consistent with the metallicity power law used to inject the samples.  

\section{Discussion}  \label{sec:discussion}
In this work, we show how known metallicity correlations with hot Jupiter occurrence will manifest in hot Jupiter occurrence rates as a function of \rbirth\ and phase-space density using GALAH, APOGEE, and simulated data from the NIHAO-UHD project \citep{Buck20b}. The correlation between metallicity and planet occurrence, particularly for giant planets, is well-established, as metal-rich stars are more likely to host giant planets \citep{Gonzalez97,Santos2004,Fischer05,Johnson10, Mortier13,Guo17,Adibekyan19,OsborneBayliss20}. We inject hot Jupiters into these samples by applying the metallicity power-law described in Equation \ref{eqn:fehPL} by \cite{Chen23}.

\subsection{High-$\alpha$ Sequence Star Sample Fraction Impacts \rbirth-planet occurrence}
When looking at how the injected planet frequency-metallicity relation impacts hot Jupiter occurrence as a function of \rbirth, we generally observe a decline in planet occurrence with increasing \rbirth\ $\geq 5$ kpc (Figure \ref{fig:feh_inj}). Declining planet frequency with increasing \rbirth\ aligns with the Galaxy's inside-out formation: the inner disk experienced more rapid and intense star formation, which leads to greater metal enrichment and results in higher planet occurrence rates. We note that recent work suggests that the radial metallicity gradient could be the consequence of an equilibrium state between metal production, accretion, and outflows, which are more efficient at larger radii \citep{Johnson24}.

Discrepancies in planet occurrence rates between GALAH, APOGEE, and the NIHAO simulation at \rbirth\ $< 5$ kpc stem from differences in sample selection. At \rbirth\ $<$ 5 kpc, the GALAH sample is biased toward a kinematically-cool population 
that is more confined to the disk plane compared to APOGEE (as shown in top right panel in Figure \ref{fig:feh_inj}). 
These differences are inherited from the metal-rich, $\alpha$-depleted stars from the low-$\alpha$-sequence shown in the [Fe/H]-[Mg/Fe] plane (Figures \ref{fig:b20dat} -- \ref{fig:apogee_hist}), more of which are observed in GALAH (95\%) compared to APOGEE and the NIHAO samples (87\% and 70\% respectively). This is a consequence of differences in survey selection function 
and does not appear to be due to an offset in survey abundances scales (e.g., \citealt{Jofre19}) since the stars in common between APOGEE and GALAH have very small offsets in \feh, \alphafe, and thus inferred \rbirth. 
In contrast to GALAH, the APOGEE survey and NIHAO simulation sample more stars away from the plane (Figures \ref{fig:b20dat} and \ref{fig:apogee_hist}) that are kinematically hotter ($|V_{\text{z}}| > 18$ km/s). These stars are more representative of the high-$\alpha$ sequence population of which APOGEE and the NIHAO samples contain more of (13\% and 30\%, respectively) compared to GALAH (5\%). These kinematic differences also correspond to mean chemical differences of 0.2--0.4 dex at a given \rbirth\ (bottom panels in Figure \ref{fig:feh_inj}), where GALAH is on average more metal-rich and $\alpha$-depleted than stars in APOGEE and the NIHAO simulation. Thus, GALAH's kinematically cooler population, which is more metal-rich is more likely to host hot Jupiters than APOGEE or the NIHAO simulation stars.

Observationally, the transition occurring at 5 kpc is reflective of where the iso-\rbirth\ lines fall in the \feh-\alphafe\ plane, as visualized in Figure \ref{fig:phs_rbirth}. Stars with \rbirth\ $= 5$ kpc contain the most equal mix of both low and high-$\alpha$ sequence stars compared to other \rbirth. As noted in Section \ref{sec:rbirth_recalc}, the absolute location of this ``turnover" is dependent on the specific \rbirth\ calibration adopted, but its physical meaning as the transition between the two chemical sequences remains consistent. The high-$\alpha$ sequence stars dominate at small \rbirth\ ($> 50\%$ at \rbirth\ bins $ < 5$ kpc in APOGEE and the NIHAO simulation). Because high-$\alpha$ sequence stars have a higher representation overall ($> 5\%$) in the APOGEE and NIHAO samples compared to GALAH, we see this transition point at 5 kpc in these two samples and not in GALAH. The magnitude limits are different between the optical GALAH and near-IR APOGEE surveys, and targeting is done in the V and H band, respectively. High-$\alpha$ sequence stars in our sample are, on average, fainter by about 0.5 magnitudes in the Gaia G-band compared to low-$\alpha$ sequence stars. Because GALAH has a brighter effective G-band cutoff than APOGEE (maximum G of $\sim$ 14 versus $\sim$ 17 for the stars we examine with cuts described in section \ref{sec:data}), this propagates to population differences between the APOGEE and GALAH samples. Therefore, APOGEE comprises a higher fraction of high-alpha disk stars (13\%) compared to GALAH (5\%).


Since GALAH predominantly observes low-$\alpha$ sequence stars, its occurrence rates reflect the expected overall decrease in planet occurrence from \rbirth\ of 0–14 kpc, consistent with the Milky Way's chemical evolution. In contrast, samples that are more heterogeneous, with a greater proportion of high-$\alpha$ sequence stars, like APOGEE and the NIHAO simulation, do not show this overall decrease. Instead, occurrence rates increase with \rbirth\ up to 5 kpc and then decrease from 5 to 0 kpc due to the dominance of high-$\alpha$ sequence stars in the inner Galaxy \citep{Buck21}. 
For a given \rbirth, APOGEE has more metal-poor stars along the track due to different density distributions of the survey that drive this relative decrease in planet occurrence compared to GALAH. 
This trend does not negate the effect of the overall radial metallicity gradient present in both disks. Rather, it reflects the relative contributions of low and high-$\alpha$ sequence stars in each sample. When limiting the APOGEE and NIHAO samples to solely the low-$\alpha$ sequence stars, we see an overall decrease in hot Jupiter occurrence rate with \rbirth\ that is now consistent with the GALAH sample. We also calculate occurrence rates separately for low- and high-$\alpha$ stars in our samples and find that, as expected from the metallicity dependence of hot Jupiter formation, the more metal-rich low-$\alpha$ population exhibits uniformly higher occurrence rates across all $\rbirth$. However, the overall trend of decreasing occurrence with increasing $\rbirth$ ultimately holds for both populations across all surveys due to the Galaxy's raidal metallicity gradient. 
These varying trends highlight the impact of survey selection effects and will be explicitly explored in future work. 


\subsection{Implications for Observed \rbirth-planet Occurrence Calculations}

Our toy model occurrence rates provide a baseline for comparison to observed occurrence rates as a function of \rbirth. Specifically, we demonstrate how planet frequency-metallicity correlations manifest in planet occurrence-\rbirth\ space. Applying the planet frequency-metallicity correlation to model \rbirth-planet occurrence rates predicts a decline of 0.1\% per kpc for \rbirth\ of 0--14 kpc in low-$\alpha$ sequence-only ($> 95\%$) samples. In mixed low and high-$\alpha$ sequence samples, occurrence rates decrease by 0.1\% per kpc for \rbirth\ $> 5$ kpc but increase by 0.2\% per kpc for \rbirth\ of 0--5 kpc. This suggests that if \rbirth\ were the primary driver of occurrence rates, observed trends in actual data with a mix of low and high-$\alpha$ sequence stars would need to show a steeper decrease than 0.1\% per kpc for \rbirth\ $> 5$ kpc and a steeper increase than 0.2\% per kpc for \rbirth\ $< 5$ kpc. Similarly, a sample with over 95\% low-$\alpha$ sequence stars would need to show a steeper decrease than 0.1\% per kpc across \rbirth\ of 0--14 kpc. Assuming no other Galactic parameter plays a role, the same or shallower observed trends to what we have modeled would indicate that metallicity plays a greater role than \rbirth\ itself in determining planet occurrence rates. 



\subsection{Phase-Space Density Classifications Approximately Map to Low and High-$\alpha$ sequences} \label{sec:simdifs}
Motivated by results from \cite{Winter20}, we calculate phase-space densities for our three samples, classifying stars into overdense and underdense environments as described in Section \ref{sec:wintermethod}. We find that on average, stars classified to be in overdense environments are more metal-rich (0.04-0.13 dex), kinematically-cool (10-22 km/s), and $\alpha$-depleted (0.01-0.1 dex) than those in underdense environments as shown in Figure \ref{fig:parentpop}. These differences echo the distinction between high and low-$\alpha$ sequence populations, and suggest that the classification is separating these populations.

We confirm that phase-space overdensities and underdensities correspond to chemically distinct Galactic populations by cross-checking stars’ $\alpha$-sequence classifications, which we explicitly tagged prior to our analysis as described in Section~\ref{sec:data}. The differences in \feh, $|V_z|$, and \alphafe\ shown in Figure~\ref{fig:parentpop} illustrate the same underlying distinctions that define the low- and high-$\alpha$ sequence populations. While most stars in each sample/parent population are low-$\alpha$ sequence stars (70-95\%), we find that in each of the three samples (NIHAO, APOGEE, and GALAH), the underdensities have an over-representation of high-$\alpha$ sequence stars by 1.2, 2.3, and 2.4$\times$, and the overdensities have an under-representation of high-$\alpha$ sequence stars by 0.93, 0.46 and 0.6$\times$. The percentage of low-$\alpha$ sequence stars present in the underdensities and overdensities reflect the distribution seen in the parent population, as they comprise the majority of the stars. So while the kinematic classification by \cite{Winter20} does not perfectly assign stars to the low and high-$\alpha$ sequence populations we designated in Section \ref{sec:data}, the underdense and overdense populations show strong correlations with high and low-$\alpha$ sequence stars, respectively. We show the exact percentages in Table \ref{tbl:percentages} with the ``boost factors" representing how over or underrepresented each low and high-$\alpha$ sequence population is within the overdensity and underdensity populations.

We note that the NIHAO simulation shows less distinction in the under/over representation of high-$\alpha$ sequence stars with boost factors closer to 1. This is also evident from the near-equal proportions of injected hot Jupiters found in overdense and underdense regions (top, right panel of Figure \ref{fig:winter}) compared to GALAH and APOGEE, where most injected hot Jupiters are found in overdense regions. However, significant chemical differences persist between the overdense and underdense populations in the NIHAO simulation (bottom row, right and left panels of Figure \ref{fig:parentpop}). This suggests that, while the low and high-$\alpha$ sequences are chemically distinct in the NIHAO simulation \citep{Buck20b}, they are less kinematically distinct compared to observed stellar samples. As a result, the phase-space density analysis is less effective at separating low and high-$\alpha$ sequence stars compared to GALAH and APOGEE data. This finding aligns with the limitations discussed in \cite{Buck20b}, where the high-$\alpha$ sequence in the NIHAO simulation is not as geometrically distinct from the low-$\alpha$ sequence, unlike in the actual Milky Way. The lack of a clear kinematic distinction between the disks in the NIHAO simulation, which the phase-density classification explicitly probes, and the simulation's physical resolution limits, also likely explain the higher variability observed in the phase-density results for the NIHAO stars (72\%) compared to the GALAH and APOGEE datasets (4\%).

\begin{deluxetable*}{cccccc}
\centering
\tabletypesize{\footnotesize}
\tablecolumns{6}
\tablewidth{0pt}
\tablecaption{Low \& High-$\alpha$ Sequence Populations in Phase-Space Over/Underdensities}
\label{tbl:percentages}
\tablehead{
    \colhead{Catalog} &
    \colhead{Parent Pop.} &
    \colhead{Underdensity Pop.} &
    \colhead{Underdensity Boost} &
    \colhead{Overdensity Pop.} &
    \colhead{Overdensity Boost} \\
    \colhead{} & 
    \colhead{\footnotesize{\% low-/high-$\alpha$}} & 
    \colhead{\footnotesize{\% low-/high-$\alpha$}} & 
    \colhead{} & 
    \colhead{\footnotesize{\% low-/high-$\alpha$}} 
    }
\startdata
GALAH & 95/5 & 88/12 & 0.9, 2.4 & 97/3 & 1, 0.6\\
APOGEE & 87/13 & 70/30 & 0.8, 2.3 & 94/6 & 1.1, 0.46 \\ 
NIHAO Sim. & 70/30 & 63/37 & 0.9, 1.2 & 72/28 & 1, 0.93
\enddata
\end{deluxetable*}


\subsection{Implications for Planet Occurrence as a Function of Phase-Density }
We examine how the planet frequency-metallicity relation manifests in hot Jupiter occurrence as a function of phase-space density. Injecting hot Jupiters using the metallicity power law reproduces the result from \cite{Winter20}, who found that more than 92\% of hot Jupiters are found around stars in apparent phase-space overdensities. \cite{Winter20} suggest that metallicity, age, and other selection effects were not confounding factors, given the similar planet host CDFs in metallicity, parallax, mass, and age space in both underdense and overdense environments. Our toy simulations replicate this result for \feh\ (middle row of Figure \ref{fig:winter}), showing similar CDFs for hot Jupiter hosts in both environments. Since our focus in this work is metallicity effects on hot Jupiters, and our most reliable measurements are for \feh\ , we only show \feh\ CDFs. We note this is an incomplete comparison to the work in \cite{Winter20}.

The similarity in \feh\ distributions for the two environments shown in the middle panel of Figure \ref{fig:winter} arises because we are comparing stars that host hot Jupiters, not the entire stellar population. Since hot Jupiter formation strongly depends on metallicity, planet hosts are dominated by metal-rich stars, resulting in similar metallicity distributions (\feh\ differences $< 0.02$ dex) across underdense and overdense environments, even though the underlying parent populations differ (\feh\ differences of 0.04-0.13 dex) as shown in the bottom two panels of Figure \ref{fig:winter} and first column of Figure \ref{fig:parentpop}. 
Thus, the metallicity dependence of hot Jupiter formation homogenizes the metallicity distributions of planet hosts across environments, masking broader differences in the parent populations. Nevertheless, the mapping of underdense and overdense regions to high and low-$\alpha$ sequence populations means hot Jupiters should predominantly be observed in the low-$\alpha$ sequence and, consequently, in overdense phase-space regions where metal-rich stars are more abundant.


This strongly implies that metallicity, rather than clustered environments, drives the hot Jupiter occurrence trends found in \cite{Winter20} and aligns with findings from \cite{Adibekyan21, Mustill22, Blaylock24}, which also challenge the conclusions of \cite{Winter20}. \cite{Adibekyan21} compared hot Jupiters in clusters (proxies for ``overdense environments") and field stars (proxies for ``underdense environments") and found no difference in orbital architectures beyond average age. Similarly, \cite{Mustill22, Blaylock24} found that older, kinematically hotter stars are less likely to host hot Jupiters. 

Our ability to reproduce \cite{Winter20}'s results using only the established planet frequency-metallicity relation suggests that the observed clustering is a byproduct of high and low-$\alpha$ sequence star properties rather than the clustering itself influencing planet formation. What appeared to be an environmental effect is more likely due to the underlying stellar properties of these stars, with metallicity being the dominant factor in shaping hot Jupiter occurrence. This underscores that, given our current understanding and the limitations of available data, host star properties, not phase-space density, are drivers of hot Jupiter formation. This also demonstrates the importance of considering the degeneracies between planet-stellar and stellar-Galactic correlations. 

\subsection{Impacts of More Physically Motivated and Rescaled \rbirth\ Calculation and NIHAO Simulation}\label{sec:rbirth_recalc}

In this work, we use Equations \ref{eqn:galahbr} and \ref{eqn:apogeebr} from \cite{Wang2024} and Mills et al. (in prep) that were based on the \rbirth\ relations found in the NIHAO simulation from \cite{Buck20b}. These calibrations linked a star’s \feh\ and \alphafe\ to inferred \rbirth\ but were not explicitly calibrated on the Sun’s theorized \rbirth\ of 5–6 kpc \citep[e.g.,][]{Minchev18,Frankel19,Lu22}. As a result, the original equations yield \rbirth\ values of 8–9 kpc for the Sun. This offset reflects the fact that the calibrations were based on the NIHAO simulation, which has a larger scale height than the Milky Way and therefore shifts the normalization of the \feh–\alphafe-\rbirth\ relation. The discrepancy highlights that \rbirth\ calculations are intrinsically relative rather than absolute \citep{Lu22b}. However, renormalization preserves the relative ordering of stars by \rbirth\ and thus the results of this work are unchanged.

Nevertheless, to explore the impact of a more physically motivated \rbirth\ scale, we recalibrate the \rbirth\ equations \ref{eqn:galahbr} and \ref{eqn:apogeebr} shifting their intercepts of 8 and 8.89 kpc downward by 2.5 and 3.39 kpc respectively to ensure the Sun’s inferred \rbirth\ is at $\sim$ 5.5 kpc (these shifted \rbirth\ tracks are what we show in Figure \ref{fig:phs_rbirth}). Given this is an absolute shift, the relative trends of planet occurrence as a function of \rbirth\ remain the same for our samples in that the highest rates of planet occurrence are towards lower \rbirth\ as shown in Figure \ref{fig:feh_inj}. For APOGEE, we find that the turnover in planet occurrence as a function of \rbirth\ lies at approximately 2 kpc, while the mean \feh\ as a function of \rbirth\ peaks around 4 kpc. As expected, this 2 kpc offset persists across both the original and recalibrated \rbirth\ scales. 
Ultimately, the recalibrated turnover still corresponds to the position where low- and high-\alphafe\ sequences are present in roughly equal numbers. We show that this more realistic anchoring of the \rbirth\ scale does not affect the interpretation of our overall results.

In parallel, we test the impact of rescaling the NIHAO simulation itself to better match the Milky Way in scale height. This rescaling is motivated by the same physical considerations that prompted recalibrating the \rbirth\ equations: to better align with the Milky Way's actual structure. However, this rescaling is performed independently of the above \rbirth\ recalibration. The intention here is to assess how this physically-motivated rescaling affects occurrence rates as a function of \rbirth\ and phase-space density derived from the NIHAO simulation. We follow \cite{Hilmi20} to scale positions and velocities uniformly by the disk scale length ratio (3.5 kpc/5.6 kpc) of the simulation from \cite{Buck20b} to the Milky Way from \cite{Bland-Hawthorn16}. This rescaling successfully aligns the simulation’s physical scale with that of the Milky Way and results in a planet occurrence turnover at \rbirth\ of 2 kpc, matching what we see with APOGEE. However, rescaling the positions and velocities reduces the phase-space density contrast between kinematically hot and cold populations. This flattening arises because the original simulation already exhibited only modest geometric differences between its low- and high-$\alpha$ sequence populations. After scaling, phase-space density classification becomes even less effective in distinguishing between them, as both populations are compressed into a narrower region of 6D space.

These tests demonstrate that our main results, derived using the original \rbirth\ calibration and unscaled simulation, are robust. The differences we discuss here reflect modest shifts or amplify existing features. These outcomes reinforce that our findings are not an artifact of absolute normalization choices, but instead reflect relative chemical and structural trends that persist even under more physically motivated assumptions.

\section{Understanding Galactic-Scale Effects on Planet Occurrence}\label{sec:conclusion}

Using a toy model approach, this work shows how metallicity correlations with hot Jupiter occurrence manifest on Galactic scales, specifically as a function of \rbirth\ and phase-space density. Low-$\alpha$ sequence-dominated ($>$ 95\%) samples like GALAH show a decrease in hot Jupiter occurrence with \rbirth, consistent with the Milky Way disk’s radial metallicity gradient, where the inner-regions of the Galaxy are more metal-rich. In contrast, mixed low and high-$\alpha$ sequence samples like APOGEE and the NIHAO simulation have occurrence rates increase up to \rbirth\ = 5 kpc due to the dominance of high-$\alpha$ sequence stars in the inner Galaxy that are on average more metal-poor than the low-$\alpha$ sequence stars in the same regions. Beyond 5 kpc, we find the same decrease in hot Jupiter occurrence as seen in GALAH that once again reflects the Milky Way’s radial metallicity gradient. We also find that when kinematically classifying stars to be in phase-space underdensities or overdensities regions following \cite{Winter20}, high-$\alpha$ sequence stars, which are on average kinematically-hotter and more metal-poor, are overrepresented in underdensities and underrepresented in overdensities. Thus, hot Jupiters are more likely to be found in overdensities, or where there are more low-$\alpha$ sequence stars, which are typically kinematically-cool and metal-rich. These findings demonstrate that the chemical and kinematic differences between low and high-$\alpha$ sequence populations, rather than intrinsic phase-space density effects, are driving observed hot Jupiter occurrence trends.

As we move towards understanding planet demographics on a Galactic scale, this study provides a crucial framework for interpreting real data. The comparisons between \rbirth, phase-space density, and metallicity trends will help clarify whether Galactic location or individual stellar properties are the dominant factors in shaping planet occurrence rates. In future work, we must also consider other factors, such as age, kinematics, $\alpha$-enrichment, and selection biases, which may play a role in shaping observed trends. Additionally, as we expand our understanding to other types of planets where metallicity correlations are less clear \citep[e.g.,][]{Petigura18}, the same careful consideration of stellar and Galactic-scale factors will be necessary to fully understand how planets form and evolve across the Milky Way.

\acknowledgements
We thank the anonymous referee for a detailed and constructive review that improved the manuscript. R.R. is supported by the NSF Graduate Research Fellowship (DGE-2236868). 

This work has made use of data from the European Space Agency (ESA) mission Gaia (https://www.cosmos.esa.int/gaia), processed by the Gaia Data Processing and Analysis Consortium (DPAC, https://www.cosmos.esa.int/web/gaia/dpac/consortium). Funding for the DPAC has been provided by national institutions, in particular the institutions participating in the Gaia Multilateral Agreement.This work made use of the Third Data Release of the GALAH Survey (Buder et al. 2021). 

The GALAH Survey is based on data acquired through the Australian Astronomical Observatory under programs: A/2013B/13 (The GALAH pilot survey); A/2014A/25, A/2015A/19, A2017A/18 (The GALAH survey phase 1); A2018A/18 (Open clusters with HERMES); A2019A/1 (Hierarchical star formation in Ori OB1); A2019A/15 (The GALAH survey phase 2); A/2015B/19, A/2016A/22, A/2016B/10, A/2017B/16, A/2018B/15 (The HERMES-TESS program); and A/2015A/3, A/2015B/1, A/2015B/19, A/2016A/22, A/2016B/12, A/2017A/14 (The HERMES K2-follow-up program). 
We acknowledge the traditional owners of the land on which the AAT stands, the Gamilaraay people, and pay our respects to elders past and present. This paper includes data that has been provided by AAO Data Central (datacentral.org.au).

Funding for the Sloan Digital Sky Survey V has been provided by the Alfred P. Sloan Foundation, the Heising-Simons Foundation, the National Science Foundation, and the Participating Institutions. SDSS acknowledges support and resources from the Center for High-Performance Computing at the University of Utah. The SDSS web site is \url{www.sdss.org}. SDSS is managed by the Astrophysical Research Consortium for the Participating Institutions of the SDSS Collaboration, including the Carnegie Institution for Science, Chilean National Time Allocation Committee (CNTAC) ratified researchers, the Gotham Participation Group, Harvard University, Heidelberg University, The Johns Hopkins University, L’Ecole polytechnique federale de Lausanne (EPFL), Leibniz-Institut fur Astrophysik Potsdam (AIP), Max-Planck-Institut fur Astronomie (MPIA Heidelberg), Max-Planck-Institut fur Extraterrestrische Physik (MPE), Nanjing University, National Astronomical Observatories of China (NAOC), New Mexico State University, The Ohio State University, Pennsylvania State University, Smithsonian Astrophysical Observatory, Space Telescope Science Institute (STScI), the Stellar Astrophysics Participation Group, Universidad Nacional Autonoma de Mexico, University of Arizona, University of Colorado Boulder, University of Illinois at Urbana-Champaign, University of Toronto, University of Utah, University of Virginia, Yale University, and Yunnan University.


\facilities{Gaia \citep{GaiaMission}, GALAH \citep{galah_mission}, APOGEE \citep{apogee}}
\software{astropy \citep{2013A&A...558A..33A,astropyii},  astroquery \citep{astroquery}, numpy \citep{numpy}, pandas \citep{reback2020pandas,mckinney-proc-scipy-2010}, topcat \citep{topcat}, matplotlib \citep{Hunter:2007}, scipy \citep{scipy}, sci-kit learn \citep{sklearn}}




\end{document}